\documentclass[epj]{webofc}
\usepackage[varg]{txfonts}   
\usepackage{hyperref}

\renewcommand*{\eqref}[1]{Eq.~(\ref{eq:#1})}

\newcommand*{\figref}[1]{Fig.~(\ref{fig:#1})}
\newcommand*{\figlab}[1]{\label{fig:#1}}

\newcommand*{\seclab}[1]{\label{sec:#1}}
\wocname{\includegraphics[width=0.25cm,clip]{logoARENA} ARENA2018}
\wocname{ARENA2018}
\woctitle{ARENA2018}
\woctitle{\includegraphics[width=0.25cm,clip]{logoARENA} ARENA2018}
\begin{document}
\title{A Surface Radio Array for the Enhancement of IceTop and its Science Prospects}
%
%

\author{\firstname{Aswathi} \lastname{Balagopal V.}\inst{1}\fnsep\thanks{\email{aswathi.balagopal@kit.edu}}
\and \firstname{Andreas} \lastname{Haungs}\inst{2}
\and \firstname{Thomas} \lastname{Huber}\inst{2}
\and \firstname{Tim} \lastname{Huege}\inst{2}
\and \firstname{Matthias} \lastname{Kleifges}\inst{3}
\and \firstname{Max} \lastname{Renschler}\inst{2}
\and \firstname{Harald} \lastname{Schieler}\inst{2}
\and \firstname{Frank} \lastname{G. Schr\"oder}\inst{1,4}
\and \firstname{Andreas} \lastname{Weindl}\inst{2}}


\institute{Institute of Experimental Particle Physics, Karlsruhe Institute of Technology (KIT), Karlsruhe, Germany
\and
           Institute for Nuclear Physics, Karlsruhe Institute of Technology (KIT), Karlsruhe, Germany
\and
	       Institute for Data Processing and Electronics, Karlsruhe Institute of Technology (KIT), Karlsruhe, Germany
\and
	  Department of Physics and Astronomy, University of Delaware, Newark, DE, USA
}

\abstract{%
Radio detection of air showers in the current era has progressed immensely to effectively extract the properties of these air showers.  
Primary cosmic rays with energies of hundreds of PeV have been successfully measured with the method of radio detection. 
There are also attempts to observe high-energy neutrinos with this technique.
Current radio experiments measuring cosmic-ray air showers mostly operate in the frequency range of 30-80 MHz. 
An optimization of the frequency band of operation can be done for maximizing the signal-to-noise ratio that can be achieved by an array of radio antennas at the South Pole, operated along with IceTop. 
Such an array can improve the reconstruction of air showers performed with IceTop. 
The prospect of using such an optimized radio array for measuring gamma rays of PeV energies from the Galactic Center is discussed.
}
\maketitle
\section{Introduction}
\label{intro}
The radio technique of air shower detection has been successfully used for the study of cosmic rays with energies well above $10^{16.5}$ eV. 
This technique can also be used to detect PeV gamma-rays, by lowering the threshold energy of detection \cite{balagopal}.
Gamma-rays of PeV energy can travel a maximum distance of $\approx$ 10 kpc before they get absorbed by the photon background from the Big-Bang (CMB). This falls within the distance between the center
of the Milky Way and the Earth ($\approx$ 8.5 kpc). Hence, we can expect to measure at least a fraction of such gamma rays approaching us from possible PeVatrons at the Galactic Center. 
A hint for the existence of such a PeVatron has already been shown by H.E.S.S. which has observed a gamma-ray spectrum with no significant cut-off at energies
up to nearly 100 TeV \cite{HESS}.

The IceCube Neutrino Observatory \cite{icecube} provides an ideal location for the search of these PeV gamma rays. Situated at the South Pole, it views the Galactic Center throughout the year, at a constant zenith angle ($\theta$) of 
$61^\circ$, which allows for a 100$\%$ duty cycle. The IceTop surface array \cite{icetop} and the future enhancement of IceTop with scintillators \cite{scint} will not be able to reconstruct air-showers from these gamma rays, 
due to their low number of particles at the observation level. Radio signals from these showers will survive and can thus be used
for their observation.  A detailed view on the entire study and the performed simulations can be found in \cite{balagopal}. Further physics potential of such a radio array is described in \cite{frankarena}.
\section{Lowering the energy threshold}
\seclab{sec-1}
One of the major challenges for the detection of air-showers with PeV energies with the radio detection technique is the relatively weak radio signal in comparison 
to the noise, at these energies. This is especially the case for the typical band of 30-80 MHz. A higher level of signal-to-noise ratio (SNR) can be 
achieved if we shift the operational frequency band to 100-190 MHz \cite{balagopal}. 
For gamma ray initiated showers with $\theta=61^\circ$, such an optimization results in a lowering of the energy threshold down to $\approx$ 1 PeV \cite{balagopal}. \figref{fig-1} shows the radio footprint of a gamma-ray shower with energy
1 PeV that falls on an antenna array with one antenna placed at each IceTop station.  For comparison, the 
footprint of a gamma-ray shower with an energy of 10 PeV is also shown.

The noise at the South Pole mainly consists of galactic and thermal components. The galactic component of the total noise decreases at high frequencies, resulting in the domination of the thermal component. 
Beyond the cross-over frequency, which is higher for lower thermal noise, we will have a constant level of noise. 
Thus, the larger the thermal noise, the higher will be the threshold energy for detection at higher frequency bands of operation. 
One can assume an antenna system with thermal noise levels of 300 K, which gives a threshold energy 
of 1.4 PeV for 100 $\%$ detection efficiency \cite{balagopal}. On the other hand, new antenna systems with low levels of thermal noise are also available. Example is SKALA with a thermal noise level of 40 K \cite{skala}. The use of such an antenna can lower the threshold 
down to 1.1 PeV (for 100 $\%$ efficiency). To determine the efficiency, $\theta = 61^\circ$ showers are simulated with random azimuth and core positions such that the core lies within 400 m from the array center. If at least 3 antennas
have a SNR $>$ 10, this shower can be detected. The resulting efficiency curves for both low and high levels of thermal noise are shown in \figref{fig-2}, left panel.
\begin{figure}[h]
\centering
\includegraphics[width=0.8\linewidth,clip]{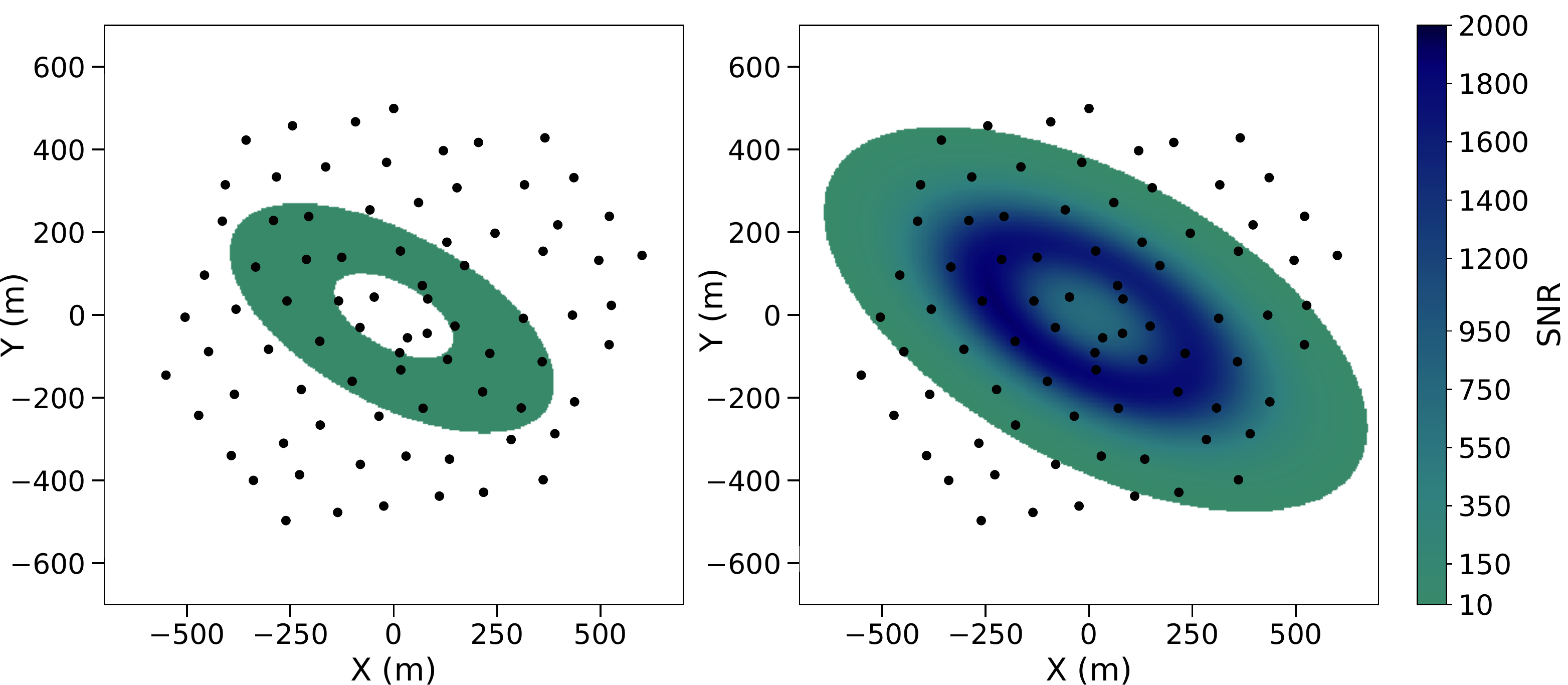}
\caption{Radio footprint at 100-190 MHz of a 1 PeV (left) and a 10 PeV (right) gamma-ray shower with $\theta = 61^\circ$ and core at (0,0). Only the antennas located at the Cherenkov ring will detect signals from the 1 PeV shower. SNR $>$ 10
is required for signal detection. Regions with SNR $<$ 10 are shown in white. The black dots show the positions of the antennas.}
\figlab{fig-1}       
\end{figure}
\vspace{-3mm}
\section{Flux estimation}
The maximum expected flux can be estimated by extrapolating the TeV gamma-ray spectrum of H.E.S.S. The gamma-rays which survive interactions
with the CMB photons can be detected by the radio array above its detection threshold. The spectrum which can be detected for two conditions of thermal noise are shown
in \figref{fig-2}, right panel. For such a spectrum with no cut-off (best-case scenario), we can expect to observe around 8 events per year for a thermal noise of 300 K and 11 events per year for a thermal noise of 40 K.
A no cut-off scenario for a point source case requires a gamma-hadron separation factor of at least 10 for a $5\sigma$ detection within 3 years \cite{balagopal}.
\begin{figure*}
\hspace{-5mm}
 \includegraphics[width=0.51\linewidth,clip]{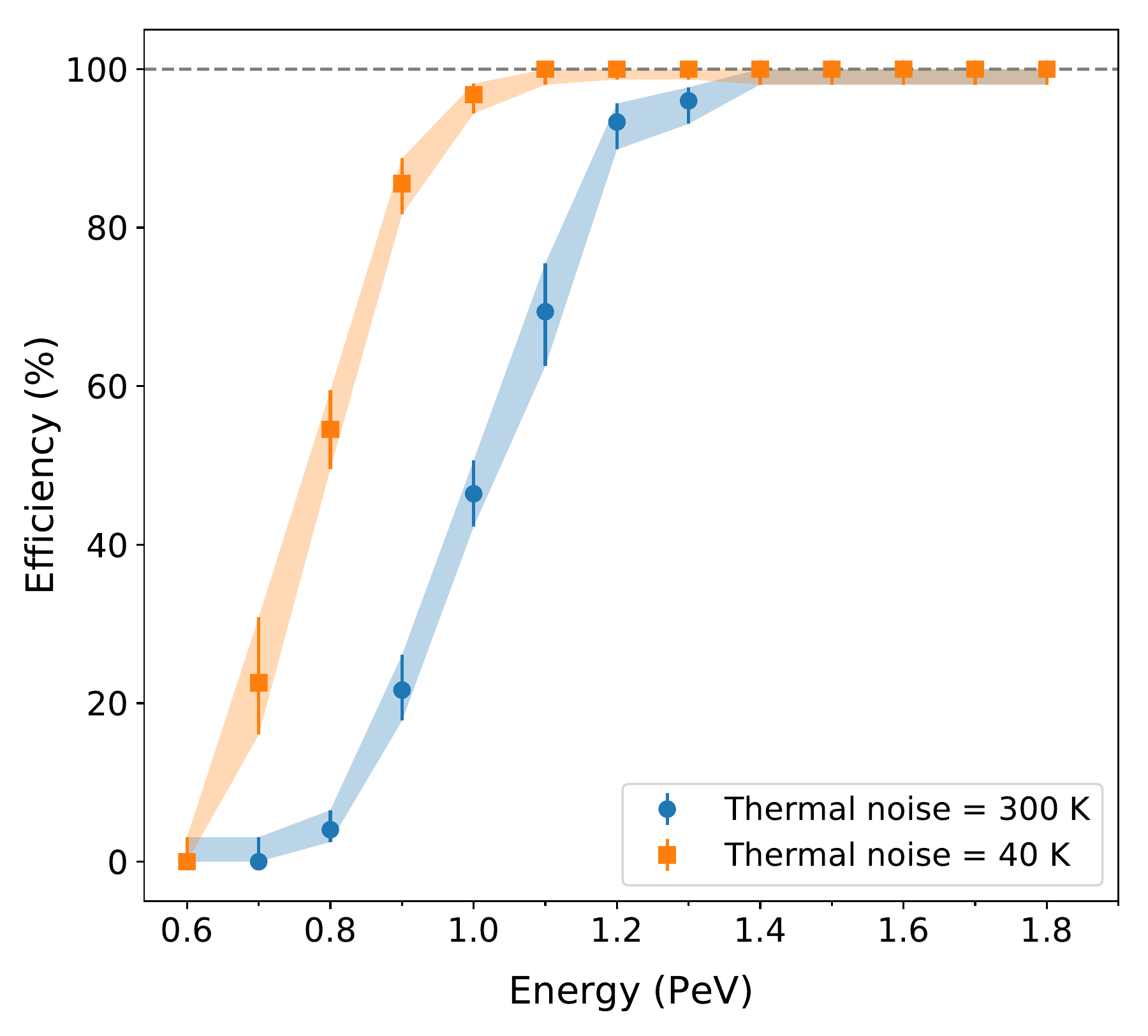}
 \hspace{-3mm}
 \includegraphics[width=0.54\linewidth,clip]{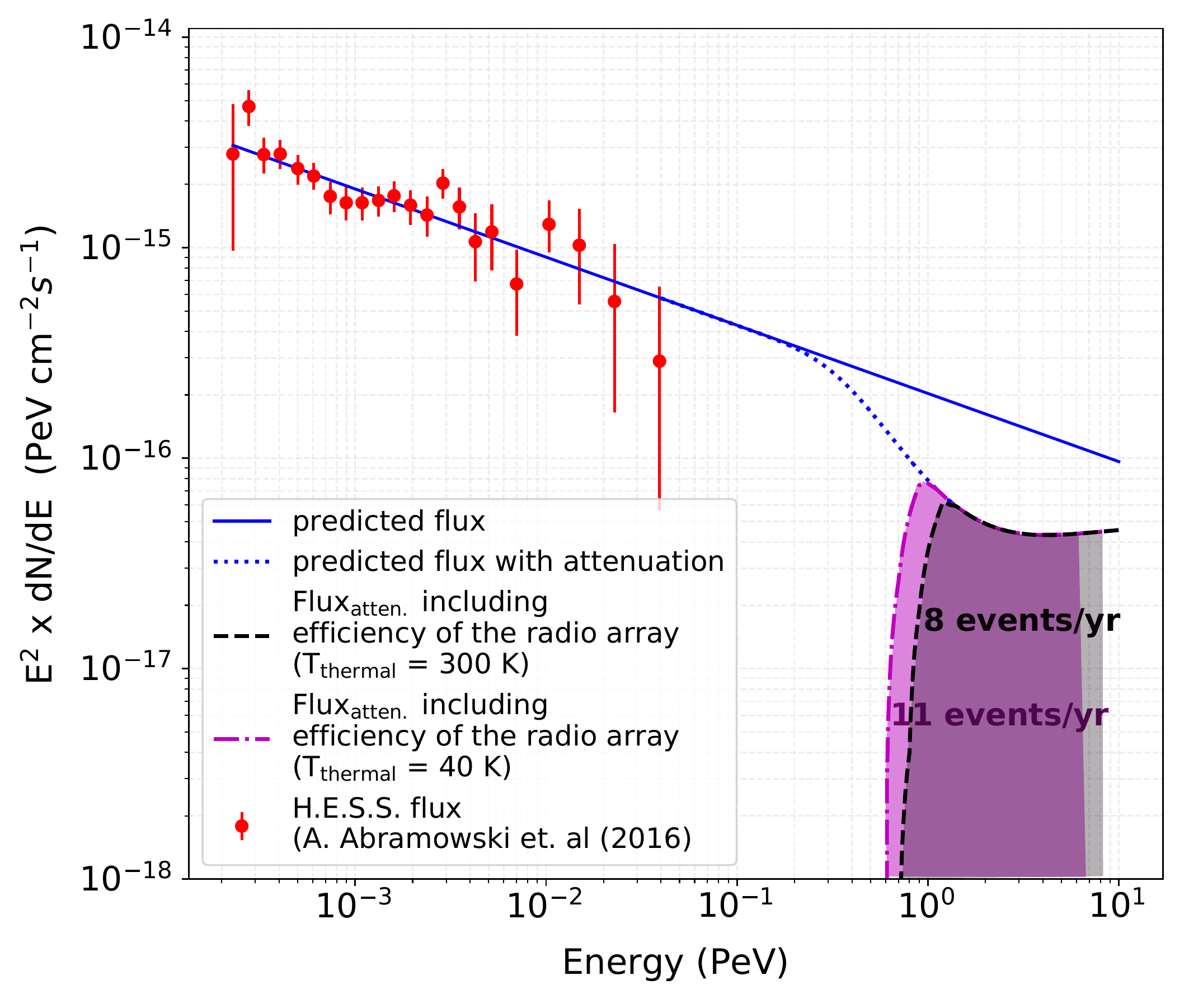}
\caption{Left: Efficiency of detection of the antenna array for high thermal noise (300 K) \cite{balagopal} and low thermal noise (40 K) conditions. Right: Expected flux (extrapolated from the H.E.S.S. spectrum \cite{HESS}) 
including the attenuation due to CMB and the efficiency of detection 
of the radio array for high \cite{balagopal} and low thermal noise.}
\figlab{fig-2}       
\end{figure*}


\section{Conclusion}
The technique of radio detection of air showers, which has so far been used for the detection of cosmic rays, can also be used to observe PeV gamma-rays approaching the Earth from the Galactic Center. This can be achieved
by deploying an antenna array operating at the optimal frequency band of 100-190 MHz at the South Pole, thereby enabling the detection of air showers at much lower energies than that of the showers detected with the radio technique so far.

\flushleft
\textbf{Acknowledgements:}
We thank the IceCube and IceCube-Gen2 collaborations for their support.

%
%
%

\end{document}